\documentclass[9pt,a4paper]{article}
\usepackage[pdftex]{graphicx}
\usepackage{ae}
\usepackage{amsmath}
\usepackage{amsbsy}
\usepackage{amsfonts}

\makeatletter

\newenvironment{figurehere}
{\def\@captype{figure}}
{}
\makeatother

\title{The $k\cdot p$ model: a short overview for beginners} 
\author{Josep Planelles\\
Dept. Química-Física i Analítica\\
Universitat Jaume I}
\date{\today}
\begin{document}
\maketitle

\section*{Motivation}
The first few contacts of a beginner with semiconductor slang can be shocking. One hears or reads that there are electrons but also holes. The beginner can figure out the hole as the vacancy of an electron, and keeps reading and/or listening. But then it appears the effective mass. It turns out that in a semiconductor, electrons behave as if their mass were different from the well-known mass of the free electron. That is, a force produces different accelerations depending on the electron being placed in a semiconductor or in a vacuum. With a little additional effort, the neophyte imagines themselves resting on the moon and being pushed  and then experiencing a different displacement than would have on earth. Right, they can imagine that the set of {\it cores} and the remaining crystal structure produce a field that intensifies or dampens the reaction to the force ... and continues reading. Difficulties grow when learning that the mass can be anisotropic. That is, depending on the force being exerted from top to bottom or from left to right, the dynamic response of the electron may be different. But one gets dumbfounded when listening that holes masses are negative. That is, should you exert a force to the right then the hole moves to the left... At this point the neophyte begins to think of Lewis Caroll and Alice in Wonderland. This doesn't look like science or rather it looks like science fiction. Well no, it is actually science. The problem with scientific dissemination is that best say nothing or say more. I will try to contextualize these definitions to make them acceptable without having to do {\it acts of faith}, which should never be done in science.

\section {Contextualizing the concepts}
The Hamiltonian eigenvalue equation of an electron in a crystal can be written as: 
\begin{equation}
\label{kp1}
\left(-\frac{\hbar ^2}{2m}\nabla ^2+V(\mathbf{r})-E_n\right)\Psi _n(\mathbf{r})=0.
\end{equation}
\noindent where the first term is the kinetic energy and  $V(\mathbf{r})$ is a periodic potential.\\

\noindent  It is useful writing the wave function in the form $\Psi _{nk}(\mathbf{r})=Ne^{i\mathbf{k}\mathbf{r}}u_{nk}(\mathbf{r})$, where $u_{nk}(\mathbf{r})$ is a periodic function. Injecting this function into equation (\ref{kp1}) leads to the so-called  $k\cdot p$ (hereafter just kp) Hamiltonian  eigenvalue equation:\footnote{For details see e.g. \cite{JP1}.} 

\begin{equation}
\label{kp2}
\left(-\frac{\hbar ^2}{2m}\nabla ^2+V(\mathbf{r})+\frac{\hbar ^2k^2}{2m}+\frac \hbar m\mathbf{k}\cdot \mathbf{p}-E_{nk}\right)u_{nk}(\mathbf{r})=0
\end{equation}

\noindent where $\mathbf{p}$ is the vectorial operator $-i\hbar \nabla$. A shorter way of writing equation (\ref{kp2}) is $(\widehat{\cal H}_{kp} -E_{nk}) |u_{nk}\rangle =0$.\\

\noindent The particular case  $\mathbf{k}=0$ of equation (\ref{kp2}) is just  equation (\ref{kp1}) for $u_{n0}(\mathbf{r})$. Solving this equation we get the eigenfunction basis set $\{u_{n0}(\mathbf{r}), n=1,2,3\dots \infty\}$ corresponding to $k=0$ (also called  $\Gamma$  point). As we move from this point ($k\neq 0$) we get another complete set of eigenfunctions $u_{nk}(\mathbf{r})$ that can eventually  be written as linear combination of the $\{u_{n0}(\mathbf{r})\}$ complete set at the $\Gamma$ point.\\

\noindent The kp Hamiltonian matrix elements $\langle u_{n0}|\widehat{\cal H}_{kp}|u_{n'0}\rangle$  read:
\begin{equation}
\label{kp3}
\langle u_{n0}|\widehat{\cal H}_{kp}|u_{n'0}\rangle= \left( E_{n'0}+\frac{\hbar^2 k^2}{2m}\right) \delta_{n,n'}+\frac{\hbar \mathbf{k}}{m} \cdot \mathbb P_{n,n'}
\end{equation}
\noindent where $\mathbb P_{n,n'}=\langle u_{n0}|\mathbf{p}|u_{n'0}\rangle$ is the so-called Kane parameter.

\subsection {The one-band model}

Should we are interested in a not overly accurate description of certain eigenvalue and the associate eigenvector, we can choose at $\mathbf{k}\neq 0$ the $\mathbf{k}=0$ function and calculate the expected value of the kp Hamiltonian with it. From the equation (\ref{kp3}), taking into account that $\mathbb P_{n,n}=\langle u_{n0}|\mathbf{p}|u_{n0}\rangle=0$ (since $\mathbf{p}$ is odd), we find that: 

\begin{equation}
\label{kp4}
E_{n}(k)= E_{n0}+\frac{\hbar^2 k^2}{2m}
\end{equation}

\noindent It turns out to be a parabolic model (the energy vs. $k$ is a parabola). The curvature (or second derivative) of $E_{n}(k)$ is precisely the inverse of the electron mass $m=1$ a.u. We may include the influence of other functions ({\it remote bands} in semiconductor slang) at second-order perturbation (we have seen that $\mathbb P_{n,n}=\langle u_{n0}|\mathbf{p}|u_{n0}\rangle=0$, therefore the first order perturbation contribution is zero): 

\begin{equation}
\label{kp5}
E_{nk}^{(2)} =-\sum_{n'} {\rm '}\frac{|\langle u_{n0}|\frac{\hbar}{m}\mathbf k\cdot\mathbf{p}|u_{n'0}\rangle|^2}{E_{n'0}-E_{n0}} 
            =\sum_{n'} {\rm '}\frac{\hbar^2 |\mathbf k \cdot \mathbb P_{n,n'}|^2}{m^2(E_{n0}-E_{n'0})}
            =\sum_{\alpha=x,y,z}\sum_{n'} {\rm '} \frac{\hbar^2 k_{\alpha}^2 \cdot |\mathbb P_{n,n'}^{\alpha}|^2}{m^2(E_{n0}-E_{n'0})}		
\end{equation}

\noindent With the inclusion of remote bands contribution, the energy finally is: 

\begin{eqnarray}
\label{kp6}
E_{nk}&=&E_{n0}+\sum_{\alpha=x,y,z} \hbar^2 k_{\alpha}^2\left\{ \frac{1}{2m}+\frac{1}{m^2}\sum_{n'} {\rm '}\frac{|\mathbb P_{n,n'}^{\alpha}|^2}{E_{n0}-E_{n'0}} \right\} \\
	  &=&E_{n0}+\sum_{\alpha=x,y,z} \frac{\hbar^2 k_{\alpha}^2}{2}\frac{1}{m^*_{\alpha}}
\end{eqnarray}
\noindent with $m^*_{\alpha}$ the so-called electron effective mass. That is, the electron is not in a semiconductor as it is in a vacuum, where it has only kinetic energy and therefore has a parabolic energy vs. the linear momentum. In a semiconductor a periodic potential is acting on the electron. However, roughly speaking, the electron behaves {\it as if}  it was in a vacuum but its mass (i.e. the inverse of the energy function curvature) has a different value that we call electron effective mass $m^*_{\alpha}$. \\ 

\noindent Moreover, if we look at the equation (\ref{kp6}) we see that the curvatures (second derivatives) can be different in different directions. This means that while the mass of the free electron is isotropic, the effective mass can be anisotropic. All the same, if we go to the literature we find out the repeated statement that the mass of the electron is highly isotropic. We will try to understand why. \\ 

\noindent In a typical semiconductor, such as $GaAs$, bands are built as linear combination of atomic orbitals. The so-called conduction band (lowest empty band) is generally a periodic linear combination in which the empty metal $s$ orbitals play the main role, while the valence bands (filled bands) involve $p$ orbitals. Other remote bands may involve $d$ orbitals, and so on. Therefore,  $p$ is the orbital closest to $s$ and therefore leads to the most important perturbation, while the other bands have smaller effects. \\ 

\noindent From what we have said, the conduction has $|S\rangle$ symmetry and then the non-zero Kane $P$ parameters with the $|P\rangle$ symmetry band  are  $P = \langle S|\hat P_x |P_x\rangle= \langle S|\hat P_y |P_y\rangle=\langle S|\hat P_z |P_z\rangle$. Therefore, the perturbation contribution to the mass is identical in all three directions (isotropic).\footnote{Many semiconductors crystallize forming Zinc-Blend structures (Symmetry $T_d$). In these structures all contributions $\langle S|\hat P_{\alpha} |D_{\beta\gamma}\rangle$ are zero, except $\langle S|\hat P_{x} |D_{yz}\rangle$ and cyclic permutations ($xyz$ has $A_1$ symmetry in $T_d$). In general, as we have said above, the actually important contribution is that of the valence  $|P\rangle$ band, and this is the reason underlying the highly isotropic character of the conduction effective mass.}\\

\noindent Let's introduce some numbers. For example, with a value of $0.303\, eV$ for the conduction band-edge energy $E_{n0}$ (that we call $E_c$), and an effective mass $m^*= 0.1\, m_0 = 0.1$ a.u., the conduction band shape around $k=0$ ($\Gamma$ point) and the free-electron band ($m=1$ a.u.) are shown together in the following figure: 

\begin{center}
\begin{figurehere}
\resizebox{0.4\columnwidth}{!}{\includegraphics{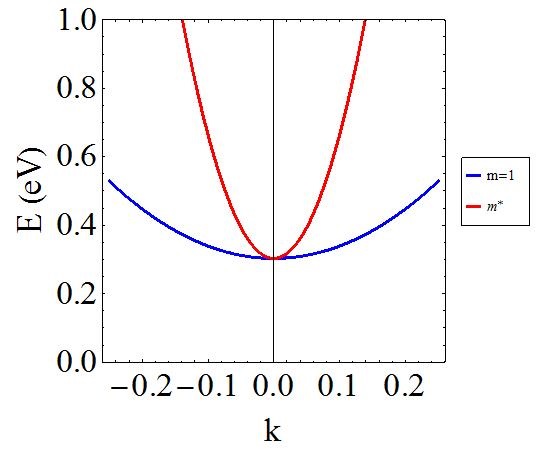}}
\end{figurehere}
\end{center}

\subsection {Two-band Model}
We may consider a conduction $|S\rangle$ and a valence $|Z\rangle$ bands. Without including the remote bands contribution, with $P = \langle S|\hat P_z |Z\rangle$, the expansion in a.u. of the Hamiltonian in this basis  $\{|S\rangle, |Z\rangle \}$ is: 

\begin{equation}
\label{kp7}
\begin{pmatrix} E_c + \frac{k^2}{2 m_e} & k P \\ k P & E_v + \frac{k^2}{2 m_h} \end{pmatrix} 
\end{equation}

\noindent For a set of values $E_c = 0.303\, eV$, $E_v=0$ (we assign zero energy to the valence band at $k=0$), $P=0$, $m_e = m_h = 1$, the form of the bands result to be: 

\begin{center}
\begin{figurehere}
\resizebox{0.35\columnwidth}{!}{\includegraphics{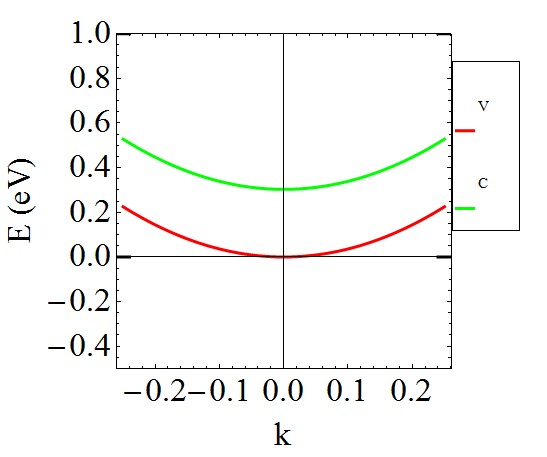}}
\end{figurehere}
\end{center}

\noindent Both, conduction and valence bands are parabolic with positive masses (curvature). Next we consider the conduction-valence interaction with e.g. the $HgTe$ Kane parameter $P=8.46$ eV$\cdot$\AA. It yields: 

\begin{center}
\begin{figurehere}
\resizebox{0.35\columnwidth}{!}{\includegraphics{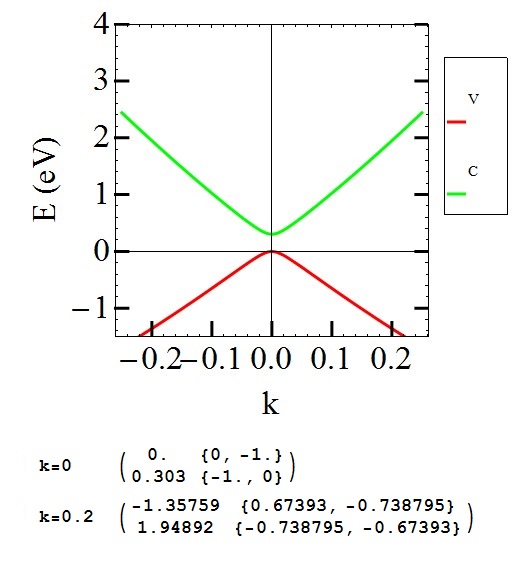}}
\end{figurehere}
\end{center}

\noindent The conduction-valence interaction turns positive masses and parabolic bands into non-parabolic ones (in this case, strongly linear) and, additionally, with an inversion of the sign in the mass of the valence band (also called the {\it holes} band). At the bottom of the figure you can see the evolution of the eigenvalues and the associated eigenvectors expanded in the basis set $\{|S\rangle, |Z\rangle \}$. For a value $k=0$, the extradiagonal terms are zero and it is clear that the conduction, corresponding to the most energetic eigenvalue (which coincides with the value $E_c = 0.303\, eV$), is the first component, while the {\it hole}, with zero energy (matching $E_v=0$) is the second one. For a value $k=0.2$ we can observe a great mixture, although keeping the largest weight that component that had all the weight at $k=0$. 

\subsection {Two-band Model with inversion}

As we already said, in general the conduction band is of $|S\rangle$ symmetry and is built out of the empty metal $s$ orbitals, whereas the valence is of $|P\rangle$ symmetry and it is basically linear combination of occupied $p$ orbitals. But there are semiconductors where this is not the case. For example, in $PbTe$, $Pb$ undergoes the so-called Lanthanides contraction while $Te$ does not. The associated volume contraction and the enormous nuclear charge of $Pb$ leads to a change in the relative energy position of the $s$ and $p$ orbitals. Then the empty band (conduction) is of $|P\rangle$ symmetry and the filled one (valence) of $|S\rangle$ symmetry. This led to the introduction of the concept of negative bandgap.\cite{Pidgeon} (See \cite{JP2} i \cite{JP3} for details). \\ 

\noindent A simulation of the inverted two-band model (just repeating the previous calculation by changing the signs of $E_c$ [$E_c = -0.303\, eV$]), results in what is shown in the following figure, plotted together with the above result for a better comparison:\footnote{Please realize that in this case (two bands with inversion) the  $|S\rangle$ band (lower energy) is fully occupied, while the $|P\rangle$ band  (higher energy) is empty.}

\begin{center}
\begin{figurehere}
\resizebox{0.6\columnwidth}{!}{\includegraphics{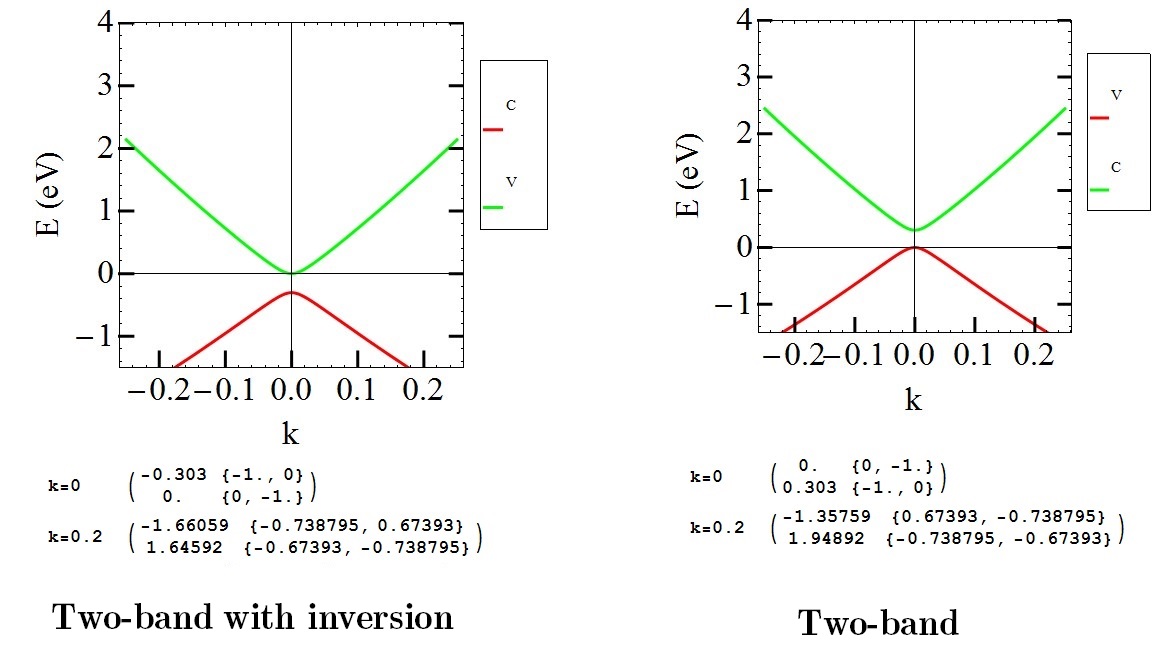}}
\end{figurehere}
\end{center}

\noindent We stress that, apart from the inversion of symmetry, in both cases the extreme of the $|P\rangle$ band, at $\Gamma=0$,  has an energy $E=0$, but while in the standard case the $|P\rangle$ band curvature is negative, in the case of $PbTe$ it is positive, with a reciprocal change in the curvature of the $|S\rangle$ band. \\ 

\noindent In this case of bands with inversion, the remote bands perturbation can generate singular bands profiles. For example, the same calculation as above but with effective masses $m_e = m_h = m_0/10$ and also a ten times smaller Kane parameter $P/10$, results in the following profile: 

\begin{center}
\begin{figurehere}
\resizebox{0.4\columnwidth}{!}{\includegraphics{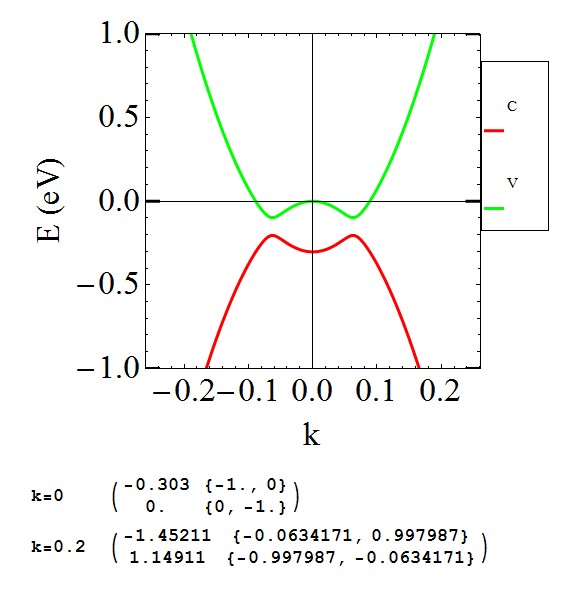}}
\end{figurehere}
\end{center}
\noindent If we look at the composition of the eigenvectors we observe a significant change in the weight of the components at $k=0$ ($\Gamma$ point) vs.  $k=0.2$.

\subsection {On the anisotropy of valence bands}
We have already said that the effective mass of the conduction band, of $|S\rangle$ symmetry, is highly isotropic. In the case of the valence band, of $|P\rangle$ symmetry, it is easy to realize that a realistic description must involve all three degenerate $p$ orbitals. Therefore, a minimally correct basis set for the valence band would be $\{|X\rangle,|Y\rangle,|Z\rangle\}$. If we take into account the spin and the often quite relevant spin-orbit coupling, the valence basis set must include six elements, which can always be chosen eigenfunctions of the total angular momentum and its $z$-component (for the  total angular momentum and its $z$-component commute with the spin-orbit operator). The corresponding quantum numbers are $J=3/2$ (four-fold degenerate), split from  $J=1/2$ (two-fold degenerate) by the spin-orbit term. Therefore, a minimally reasonable description of the top of the valence band must involve the four-fold degenerate states $J=3/2$ ($J_z=\pm 3/2$ called  heavy hole HH, $J_z=\pm 1/2$, called light hole LH). The corresponding base is: 
\begin{equation}
\begin{array}{ll}
\label{kp8}
|3/2,3/2\rangle =-\frac{1}{\sqrt{2}} |(X + i \, Y) \uparrow\rangle & |3/2,-3/2\rangle =\frac{1}{\sqrt{2}} |(X -i \, Y) \downarrow\rangle \cr
\cr
|3/2,1/2\rangle =\sqrt{\frac{2}{3}} |Z \uparrow\rangle - \frac{1}{\sqrt{6}}  |(X + i \, Y) \downarrow\rangle & 
                                    |3/2,-1/2\rangle =\sqrt{\frac{2}{3}} |Z \downarrow\rangle + \frac{1}{\sqrt{6}}  |(X - i \, Y) \uparrow\rangle  
\end{array}         
\end{equation}

\noindent Let's consider for example the interaction of HH (3/2, 3/2) with the closest state (of  $|S\rangle$ symmetry). The contribution to the effective mass of $k_{\alpha}$ involves the term $\langle S|\hat P_{\alpha} | -\frac{1}{\sqrt{2}} (X+i Y)\rangle$, so the contribution $\alpha=z$  is zero, while the two (equivalent) contributions $\alpha =x,y$ are not. We can do the same reasoning if we disregard the spin. In this case we consider, for example, the state $|Z\rangle$. The contribution to the effective mass of $k_{\alpha}$ to this state involves the term $\langle S|\hat P_{\alpha} |Z\rangle$. Therefore it is zero if $\alpha =x,y$, whereas it is not if $\alpha =z$. Therefore, the effective masses in the $z$ direction are different from the masses {\it in-plane} $m_{xy}$. We can therefore easily understand why the {\it holes} effective masses are highly anisotropic $m_z \neq m_{xy}$.

\section {Last comment}
Of course,  there is a lot more in kp than this short overview. The aim here has been to show that, indeed, coming to study semiconductors is, in a way, like Alice crossing the gate and entering wonderland, where the empty place left by an electron comes to life in the form of {\it hole}, this {\it hole} behaving quite strange: when  you push it, instead of separating from you, it turns against you. And you have to pay attention if you apply a force: depending on whether the direction is from top to bottom or from left to right the answer can be very different ....

\section*{Appendix }
We include here the Mathematica(R) code employed to generate all the figures and eigenvectors: \\

\begin{center}
\begin{figurehere}
\resizebox{\columnwidth}{!}{\includegraphics{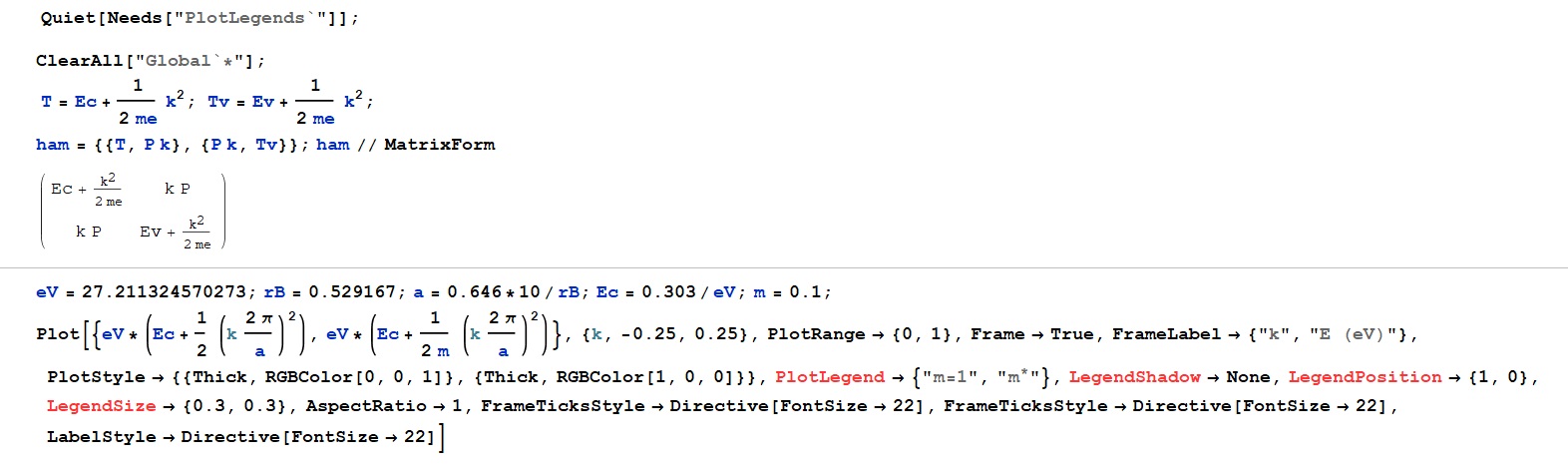}}
\end{figurehere}
\end{center}

\begin{center}
\begin{figurehere}
\resizebox{0.25\columnwidth}{!}{\includegraphics{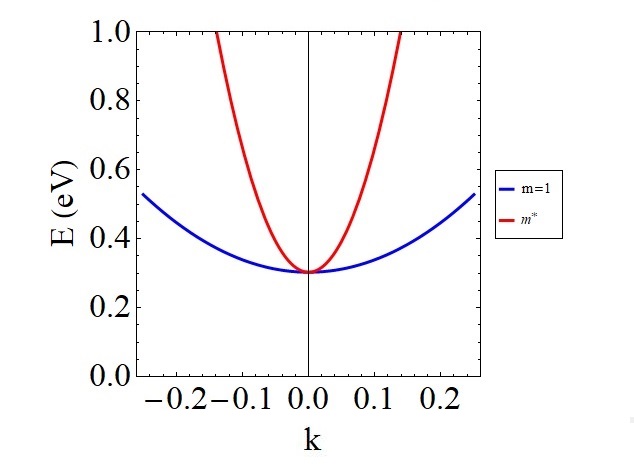}}
\end{figurehere}
\end{center}

\begin{center}
\begin{figurehere}
\resizebox{\columnwidth}{!}{\includegraphics{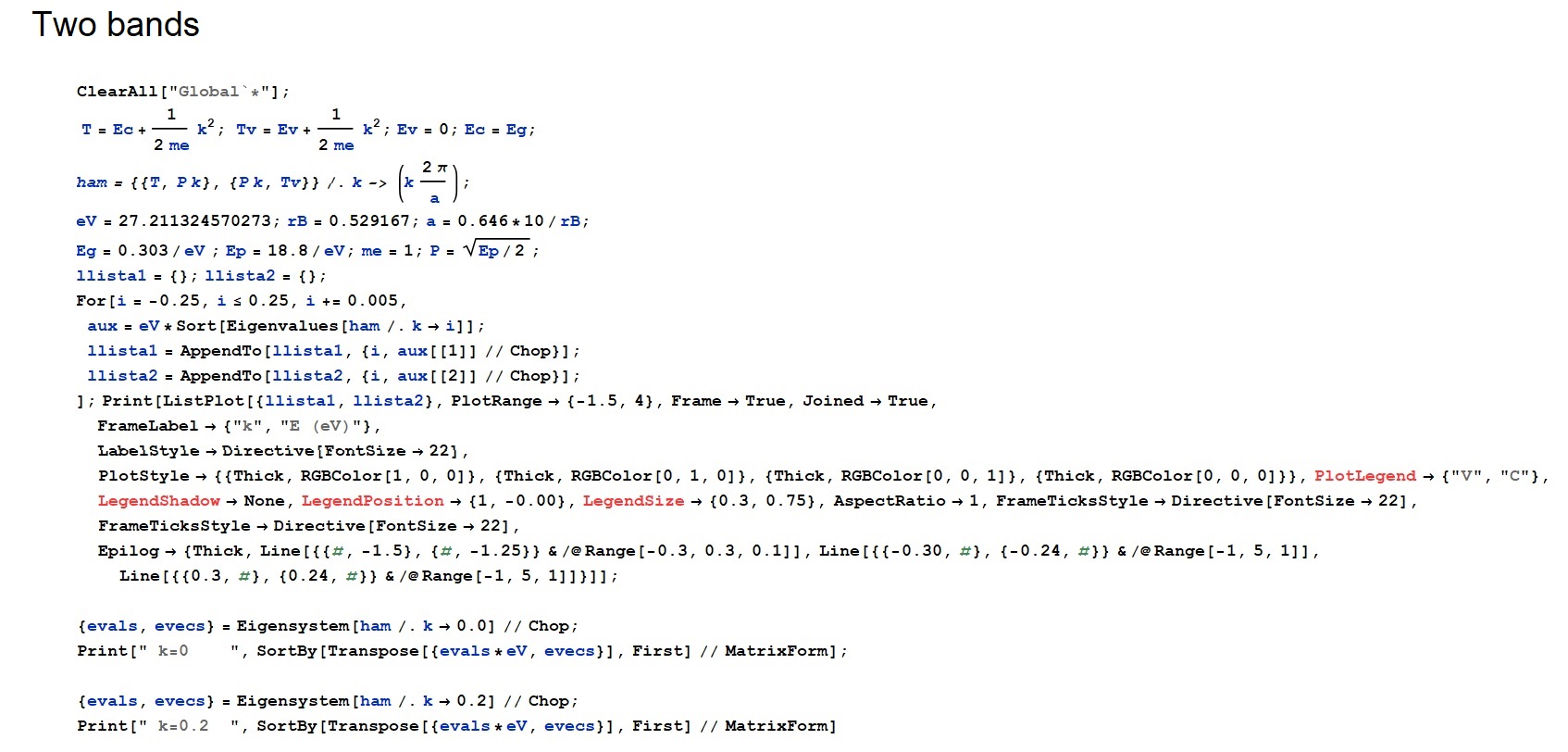}}
\end{figurehere}
\end{center}
\begin{center}
\begin{figurehere}
\resizebox{0.5\columnwidth}{!}{\includegraphics{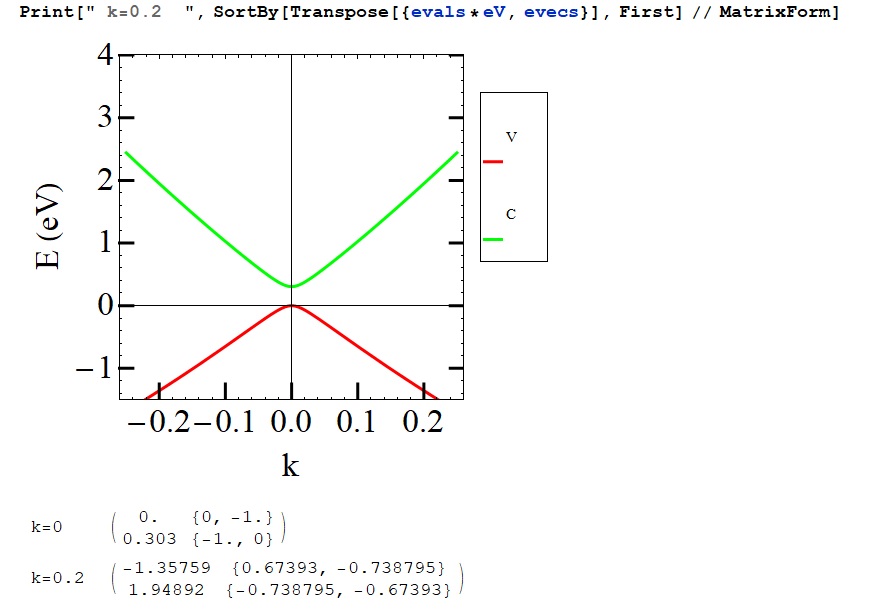}}
\end{figurehere}
\end{center}
\begin{center}
\begin{figurehere}
\resizebox{\columnwidth}{!}{\includegraphics{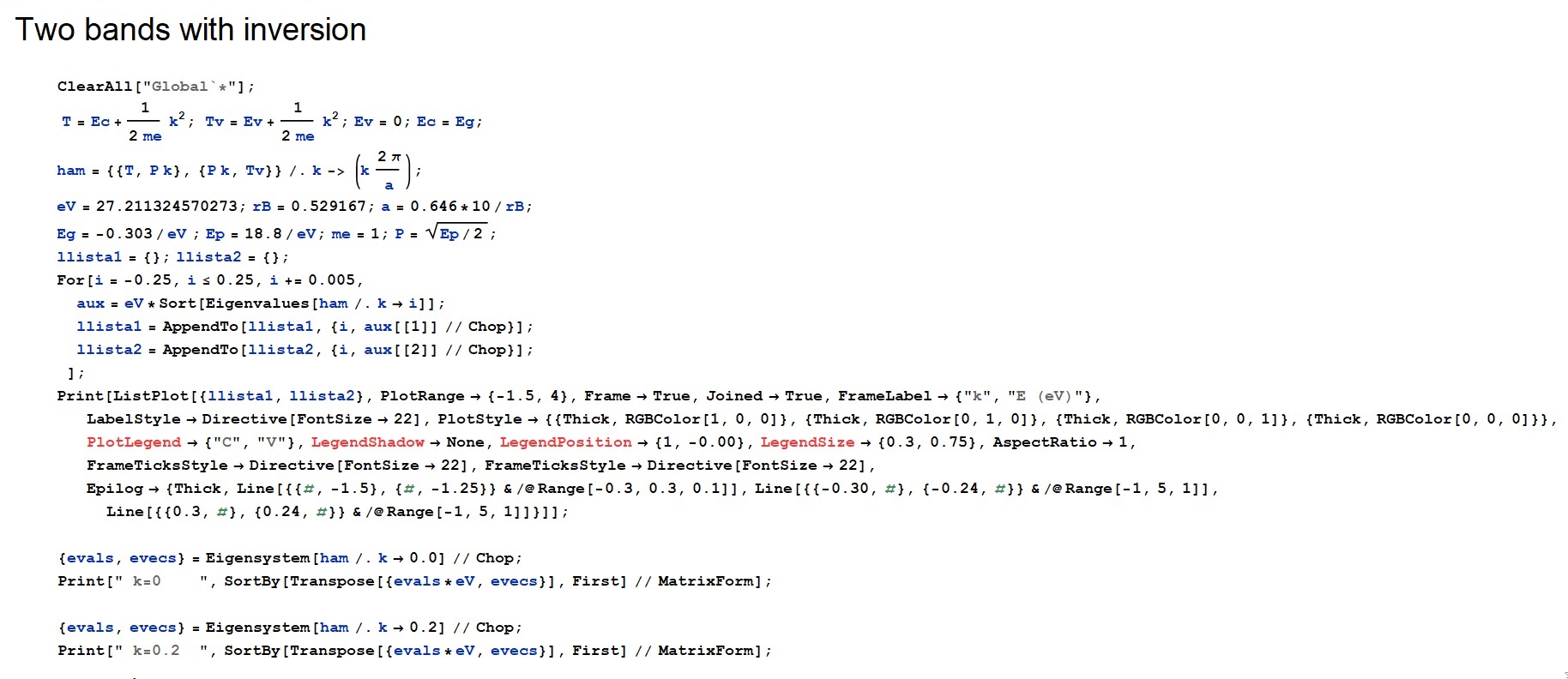}}
\end{figurehere}
\end{center}
\begin{center}
\begin{figurehere}
\resizebox{0.35\columnwidth}{!}{\includegraphics{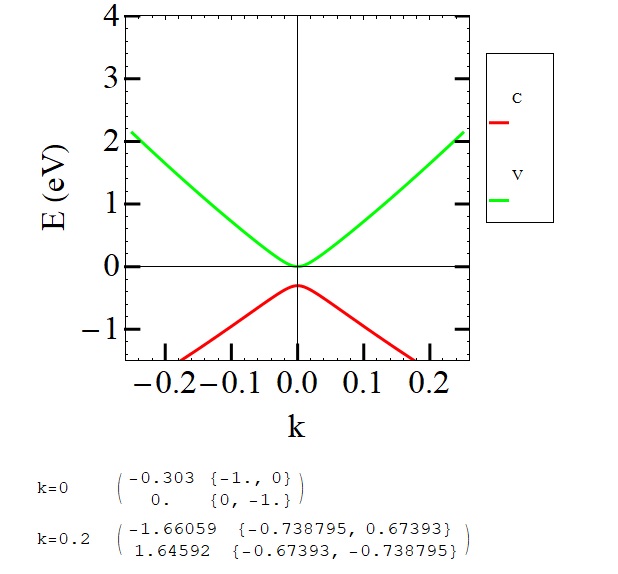}}
\end{figurehere}
\end{center}
\begin{center}
\begin{figurehere}
\resizebox{\columnwidth}{!}{\includegraphics{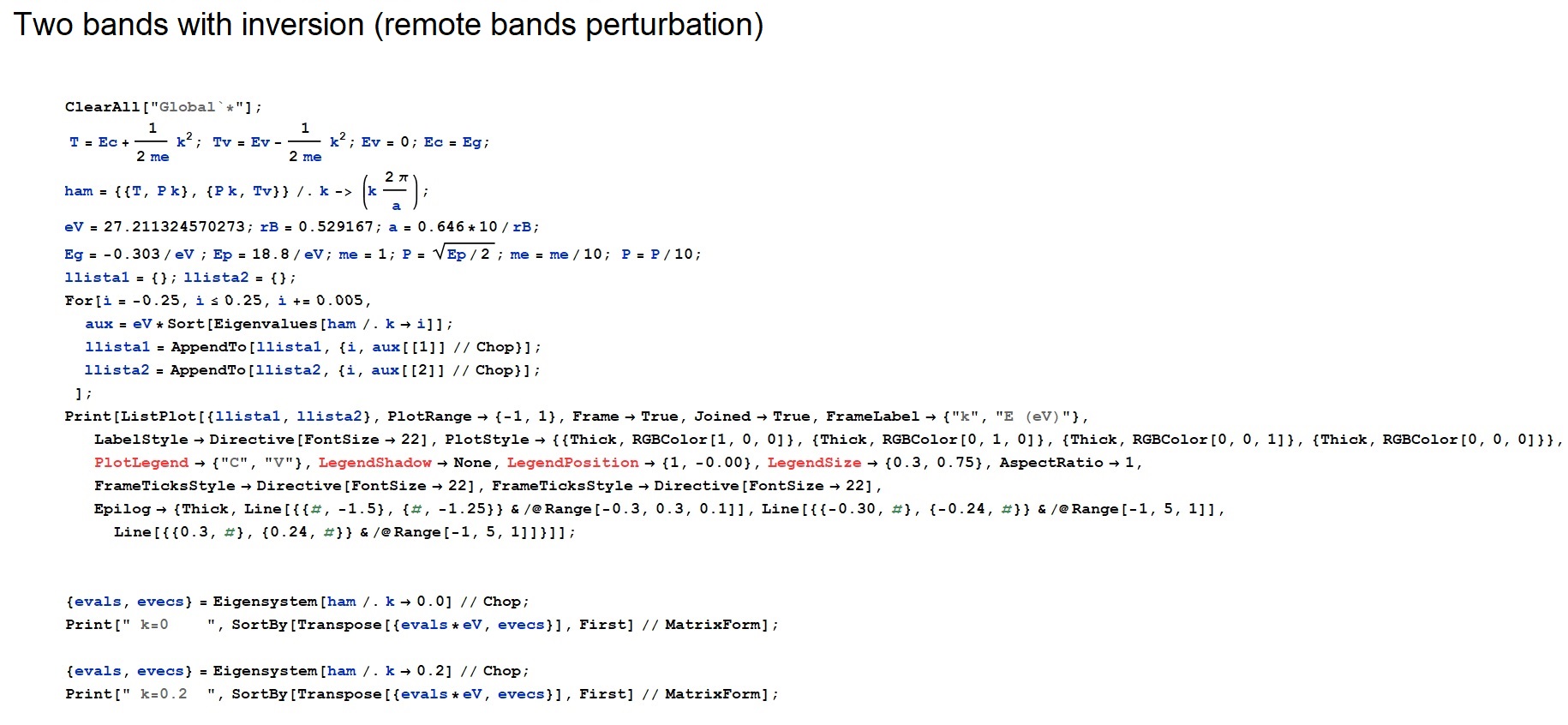}}
\end{figurehere}
\end{center}
\begin{center}
\begin{figurehere}
\resizebox{0.35\columnwidth}{!}{\includegraphics{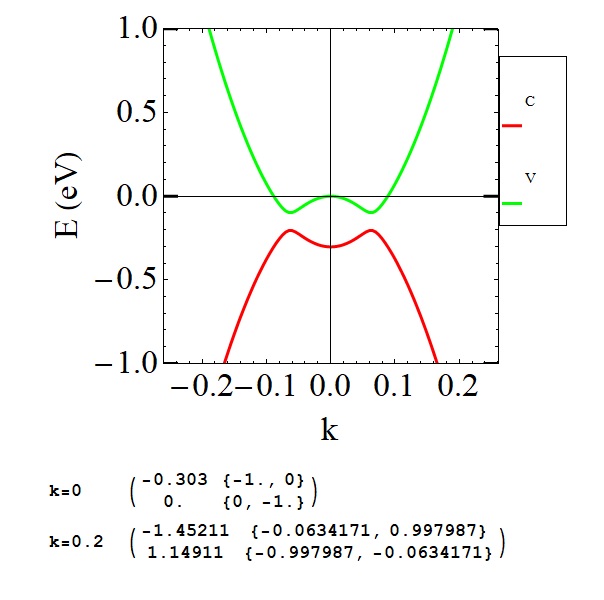}}
\end{figurehere}
\end{center}

\end{document}